\documentclass[aps,pre,twocolumn,groupedaddress,floatfix,superscriptaddress]{revtex4}

\usepackage{dcolumn}
\usepackage{amsmath}
\usepackage{amssymb}
\usepackage{lipsum}

\usepackage{graphicx}
\usepackage{bm}
\usepackage[T1]{fontenc}
\usepackage{color}
\usepackage{url}
\usepackage[bf]{subfigure}
\usepackage{rotating}
\usepackage{soul}
\usepackage{scalefnt}
\usepackage[hidelinks ]{hyperref}
\usepackage{scalefnt}

\usepackage{perpage}
\MakePerPage{footnote}

\usepackage{color}
\definecolor{redcolor}{rgb}{1.0,0.,0.}

\begin{document}
\title{Connecting Network Science and Information Theory}

\author{Henrique F. de Arruda}
\email{h.f.arruda@gmail.com}
\affiliation{Institute of Mathematics and Computer Science, University of S\~ao Paulo, S\~ao Carlos, SP, Brazil.}
\author{Filipi N. Silva}
\affiliation{S\~ao Carlos Institute of Physics, University of S\~ao Paulo, S\~ao Carlos, SP, Brazil}
\author{Cesar H. Comin}
\affiliation{S\~ao Carlos Institute of Physics, University of S\~ao Paulo, S\~ao Carlos, SP, Brazil}
\author{Diego R. Amancio}
\affiliation{Institute of Mathematics and Computer Science, University of S\~ao Paulo, S\~ao Carlos, SP, Brazil.}
\author{Luciano da F. Costa}
\affiliation{S\~ao Carlos Institute of Physics, University of S\~ao Paulo, S\~ao Carlos, SP, Brazil}

\begin{abstract}

A framework integrating information theory and network science is proposed, giving rise to a potentially new area.  By incorporating and integrating concepts such as complexity, coding, topological projections and network dynamics, the proposed network-based framework paves the way not only to extending traditional information science, but also to modeling, characterizing and analyzing a broad class of real-world problems, from language communication to DNA coding.  Basically, an original network is supposed to be transmitted, with or without compaction, through a sequence of symbols or time-series obtained by sampling its topology by some network dynamics, such as random walks.  We show that the degree of compression is ultimately related to the ability to predict the frequency of symbols based on the topology of the original network and the adopted dynamics.  The potential of the proposed approach is illustrated with respect to the efficiency of transmitting several types of topologies by using a variety of random walks.  Several interesting results are obtained, including the behavior of the Barab\'asi-Albert model oscillating between high and low performance depending on the considered dynamics, and the distinct performances obtained for two geographical models. 

\end{abstract}

\maketitle

\setcounter{secnumdepth}{1}

\section{Introduction}

A great deal of efforts in science and technology has been focused on the study of information theory~\cite{Cover:2006:EIT:1146355} and network science~\cite{Newman:2010:NI:1809753,barabasi2016network}, two seemingly independent realms.   In information theory, basically, probabilities are assigned to symbols and used to derive important results, such as minimum bandwidth and minimal sampling rates.  On the other hand, in \emph{network science}~\cite{Newman:2010:NI:1809753}, focus is given to understanding the intricate topology of complex networks, and its relationship with various types of dynamics.  Interestingly, these two different perspectives --- broadly related to time series compaction and studies of topology/dynamics complexity --- can be shown to ultimately be intertwined and complementary one another. For instance, information theory has been used to define causal relationships between nodes~\cite{borge2016dynamics,sun2015causal}, characterize networks according to their compressibility~\cite{ahnert2014generalised}, define topological similarity~\cite{de2016spectral}, map time-series to networks~\cite{lacasa2014network}, quantify the diversity of ecological networks~\cite{ulanowicz20119}, and characterize network dynamics~\cite{andjelkovic2015hidden}. 

\begin{figure}
  \centering    \includegraphics[width=0.40\textwidth]{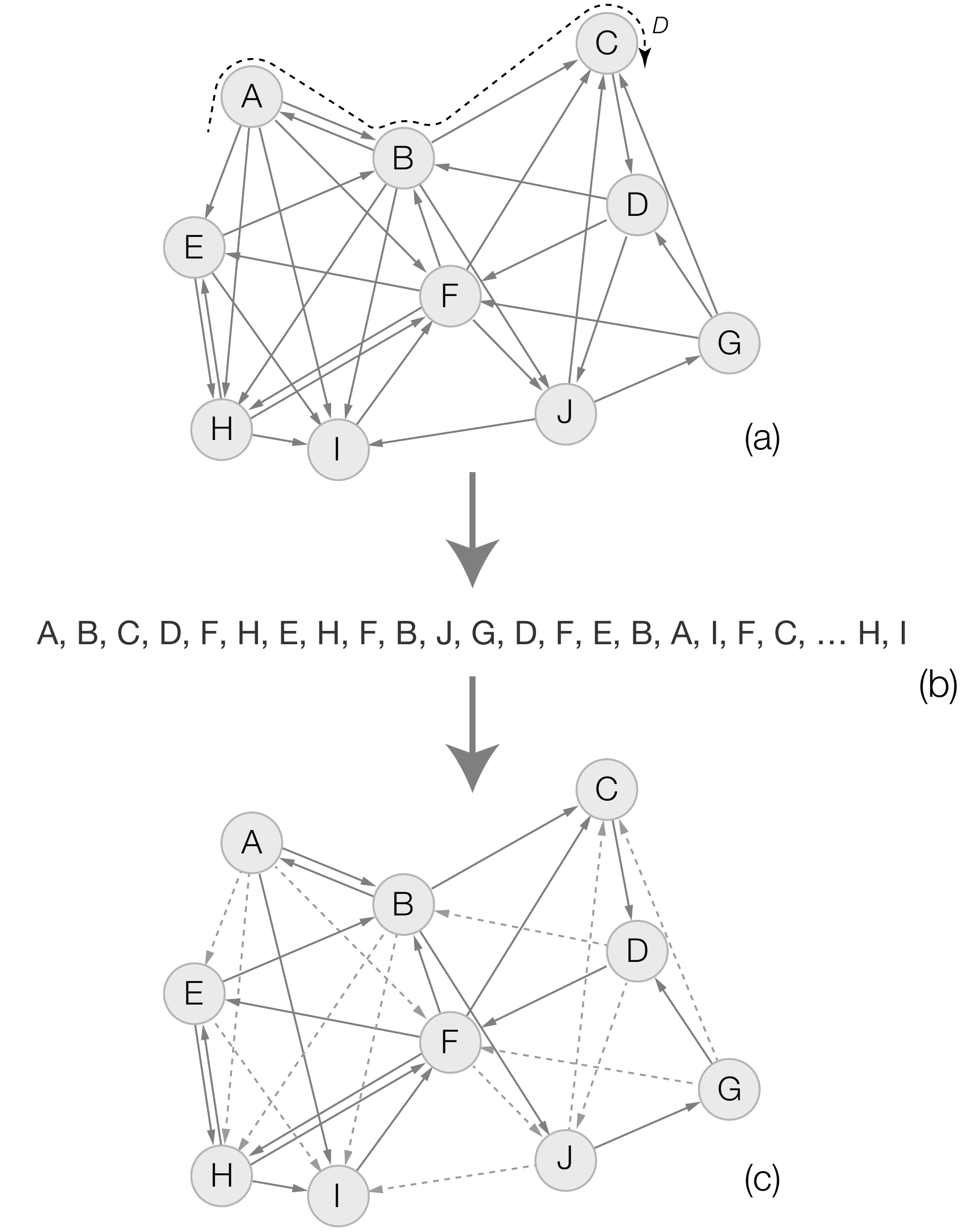}
  \caption{Overall framework underlying the integration between network science and information theory. A network (a) is sampled by a given dynamics $D$ and a respective time series corresponding to a sequence of symbols (b) is obtained and transmitted.  The receiver of the latter can then try to recover the original network (c). If desired, the transmission can be optimized, e.g. by using compaction.}\label{f:intNetTrans}
\end{figure}

A systematic integration of information theory and network science so as to provide a unified scientific approach, constitutes the main purpose of the current article.  The basic idea is to understand a sequence of symbols or time series as a projection of an original network, e.g. obtained by some sampling dynamics (such as random walks), transmitted, and then reconstructed with some accuracy.  This basic framework is illustrated in Fig.~\ref{f:intNetTrans}.

Underlying such an approach is the hypothesis that every time series or sequence of symbols is produced by some discrete system, which can be represented as a complex network.  So, these generated series and sequences inherit, to a great extent, the properties of the generating networks.  The second underlying hypothesis of the present work is that interaction between such complex network systems has to proceed through communication channels, which necessarily have limited bandwidth and/or are noisy.  So, it becomes important to devise and consider methods for more effective/robust transmission of the networks, such as using compression, which is the main original motivation behind information theory.   At the same time, these series of symbols are a byproduct of the interaction between the topology and the dynamics unfolding in the original network.   Therefore, the proposed approach transfer the design of effective communication methodologies from looking only at the time series to the network level.

\begin{figure}
  \centering
    \includegraphics[width=0.45\textwidth]{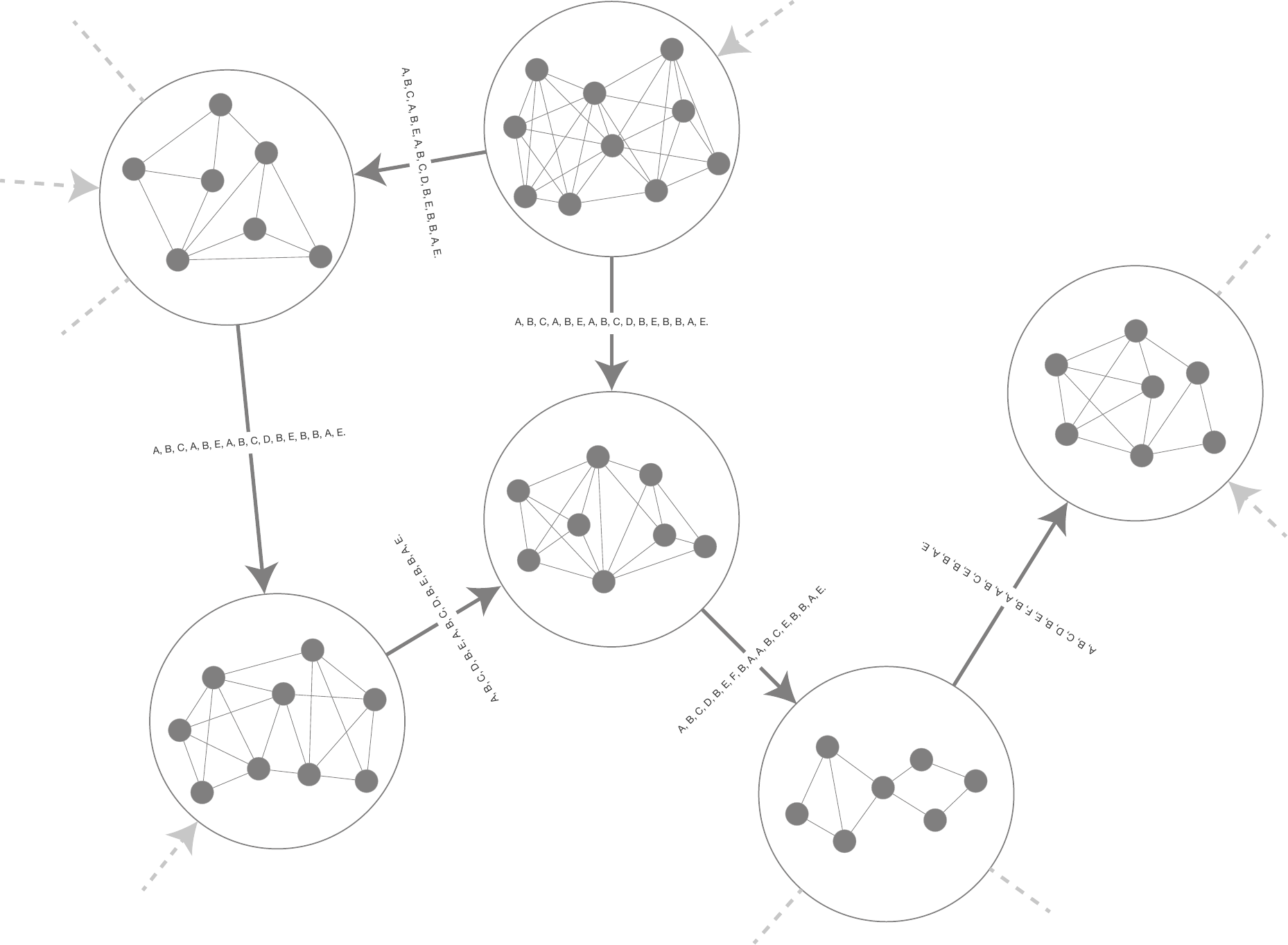}
  \caption{The integration of network science and information theory can be used to address systems involving several interactions of the type represented in Fig.~\ref{f:intNetTrans}, with applications in ares such as opinion spreading.}\label{f:intMultNet}
\end{figure}

The potential of such an integration is broad, as several important problems can be naturally conceptualized and represented according to the proposed framework.  Possible applications include, but are not limited to: the transformation from thoughts into language, encompassing the whole of literature in the process, the creation and reception of artistic pieces such as in music, routing in transportation and computing systems, optimal data allocation in distributed computing, teaching and planning of syllabuses, economics and financial indices, the codification of proteins and genes into linear chains, and even WWW surfing and the flow of consciousness.  An interesting aspect shared by all such cases is that the linearization of a higher dimensional structure (a network) derives from imposed constraints such as finite bandwidth channel, storage systems, etc.  By the way, intermediate representations with dimensions higher than one (time series) are also possible and naturally incorporated in the proposed theory.  In addition, the efficiency of coding by projections and respective transmission are related to the \emph{complexity} of the original network, therefore emphasizing another critical issue shared by information theory and network science.  Though we have so far been restricted to point-to-point communication, the proposed framework also extends naturally to larger systems involving many such pairwise interactions, such as illustrated in Fig.~\ref{f:intMultNet}.  Several real-world systems can be represented and studied by using such a framework, such as opinion spreading and evolution of scientific ideas.  In such cases, each of the large nodes in Fig.~\ref{f:intMultNet} would correspond to an agent transmitting its beliefs, originally represented by the networks inside the nodes, through time series along a network with a particular topology.

Optimization of the transmission can be achieved by compacting the time-series.  In information theory, this is typically achieved by using the frequency of symbols as a means to derive optimal code words. For instance, Huffman coding~\cite{Cormen:2001:IA:580470} provides a means to achieve lossless, optimal symbol-by-symbol coding of time series, considering the probability of each symbol.  In the here proposed framework, optimization implies in accurate prediction of symbols by considering the topology of the network and the probing dynamics, instead of only the sequence of symbols.   For instance, in the case of symbols taken individually from an undirected graph by using a traditional random walk, it is known that the frequency of each symbol can be perfectly predicted from the respective node degrees~\cite{da2007correlations}.  The precision of such a prediction can be expressed in terms of the Pearson correlation coefficient between the frequency of visits and degrees (or other topological property).  Because of its ability to characterize how much the dynamics is affected by the topological features of the network, such a measurement is henceforth called \emph{steering coefficient} of the topology over the dynamics, being henceforth represented as $S$.  All in all, the efficiency of coding given a specific network and dynamics will probably depend on the value of the steering coefficient, which can vary largely among different network topologies and dynamics.  

The present work illustrates and explores the potential of the proposed framework with respect to synthetic networks, which allows the consideration of several topologies and different sampling dynamics.  Through such a procedure, we can investigate interesting questions such as: (i) how does the ability to predict the frequency of symbols impact the transmission?; (ii) how do the topological features of different graph models affect the performance?; (iii) how do the considered types of dynamics compare regarding the time of exploration?.  These questions are tackled by considering as objective the  recovery of the original graph with a given accuracy.

\section{Methodology} 
   
\subsection{Adopted dynamics} 
In order to investigate the proposed framework, we adopted four random walk dynamics and used them to generate respective sequences of symbols. The considered dynamics are: the random walk (RW)~\cite{lovasz:1993}, a variation in which the transition probabilities are biased toward nodes with higher degree (RWD)~\cite{bonaventura2014characteristic}, another variation in which the inverse of the node degree is considered (RWID)~\cite{bonaventura2014characteristic}, and the true self-avoiding random walk (TSAW)~\cite{amit1983asymptotic,kim2016network}.

In a traditional RW dynamics, the next node to be taken by an agent is selected uniformly among its neighboring nodes. In a degree-biased random walk, the probability $p_{ij}$ that the agent goes from node $i$ to node $j$ depends on the degree of each neighboring node. Here we consider a dependence of the form

\begin{equation}
	p_{ij} = \frac{k_j^{\alpha}}{\sum\limits_{l\,\in\,\Gamma_i}k_l^{\alpha}},
\end{equation}
where $\Gamma_i$ is the set of nodes connected to node $i$. When $\alpha=1$ we have the RWD dynamics, while $\alpha=-1$ results in the RWID case. An interesting property of the RW dynamics is that, on undirected networks, as the generated sequence increases, the frequencies of visits to nodes as inferred from the current number of times each node has been visited, become directly proportional to the node degree~\cite{da2007correlations}. For the degree-biased case, the steady state probabilities of a degree-biased random walk can be written as~\cite{PhysRevE.78.065102}
\begin{equation}
P_i = \frac{\sum\limits_{j\in\Gamma_i}k_j^{\alpha} k_i^{\alpha}}{\sum\limits_{h}\sum\limits_{l\in\Gamma_h}k_l^{\alpha} k_h^{\alpha}}.\label{eq:probSteady}
\end{equation}
In case the degrees of the neighbors of node $i$ can be approximated as the average degree of the network (which happens, for instance, for narrow degree distributions), the numerator of Equation~\ref{eq:probSteady} can be written as $\sum\limits_{j\in\Gamma_i}k_j^{\alpha}\approx \bar{k}^{\alpha}k_i$, leading to

\begin{equation}
P_i \approx \frac{\bar{k}^{\alpha} k_i k_i^{\alpha}}{\bar{k}^{\alpha}\sum\limits_{h}k_h k_h^{\alpha}} = \frac{k_i^{\alpha+1}}{N\bar{k^{\alpha+1}}}\label{eq:longTermProbApprox}.
\end{equation}
This means that the probabilities become related to the degree taken to $\alpha + 1$.  

In the TSAW, the memory of the path already taken by the agent is kept and considered for determining its next step. Here, we opted to consider the frequency of edges in contrast to the frequency of nodes~\cite{kim2016network}. In this way, edges already visited many times by the agent are avoided.  Thus, the probability $p_{ij}$ that the agent moves through a certain edge $e$ at its immediate neighborhood $Ne$ is 
\begin{equation}
	p_{ij} = \frac{\gamma^{-f_e}}{\sum\limits_{j\,\in\,Ne}{\gamma^{-f_j}}},
\end{equation}
where $\gamma$ is a parameter of the dynamics and $f_e$ is the frequency of visits to edge $e$. Note that self-avoiding behaviour is achieved for $\gamma>1$. The TSAW dynamics can be expected to be usually faster to cover a network compared to the RW since unvisited connections are prioritized~\cite{kim2016network}. For large number of iterations, the dynamics tends to behave like a diffusion,  i.e. similarly to the RW dynamics~\cite{amit1983asymptotic,amit1983asymptotic}. In the analysis we set $\gamma=2$.

\subsection{Complex network models}

Six network models were used to investigate the proposed framework, namely the Erd\H{o}s-R\'enyi (ER)~\cite{erdos:1960}, Barab\'asi-Albert (BA)~\cite{Barabasi99:Science}, Watts-Strogatz (WS)~\cite{Watts98:Nature}, Waxman (WAX)~\cite{waxman1988routing}, random geometric (GEO)~\cite{dall2002random} and Knitted (KN)~\cite{costa2007knitted} models.  Given its nature, the latter model is used only on experiments involving directed networks. The ER model generates small-world networks having a binomial degree distribution~\cite{Newman:2010:NI:1809753}, which means that all nodes in the network have similar degree. In contrast, networks generated by the BA model have a power-law degree distribution~\cite{Barabasi99:Science}, implying that a few nodes, called hubs~\cite{barabasi2016network}, possess large degree while most of the nodes in the network have low degree. The WS model can be used to generate networks having the small-world property while also possessing large clustering coefficient values~\cite{Watts98:Nature}. We adopt a variation of this model where instead of a ring one starts with a lattice network and edges are rewired with probability $p$. We consider two rewiring probability values for the WS model: $p=0.01$ (WS1) and $p=0.005$ (WS2). In the GEO model, nodes are randomly placed, with uniform probability, in a two-dimensional space and pairs of nodes are connected if their distance is smaller than a given value. The WAX model begins with the same node placement procedure as in the GEO model, but pairs of nodes are connected according to a probability that decays exponentially with the distance between the nodes. Networks generated by the GEO and WAX models tend to have large diameter. KN networks are formed by treading paths. Initially, a set of unconnected nodes is created. Then, a sequence of distinct nodes is randomly selected and visited until a stop criterium is reached. Adjacent nodes in the sequence are connected through respective directed links. The process can be repeated many times until a desired average degree is reached. KN networks are peculiar among the considered models because it can be understood as being generated by a random walk. Such networks can be used for modeling co-occurrence networks in texts and other real-world situations involving sequential, uninterrupted visits to nodes~\cite{Cohen2005,Liu2013,1742-5468-2015-3-P03005,0295-5075-98-1-18002}.

The network models presented above, with the exception of KN networks, correspond to undirected networks. In this study we also considered directed networks. In order to convert the original network into a directed one, we assigned directions to the edges. First, we defined a parameter $r$, called reciprocity, which is the probability of an edge to have both directions (i.e., be reciprocal). For each original edge, a random number $n$ in the range $[0,1]$ was generated with uniform probability. If $n \leq r$, the original edge was split into two edges (in and out); otherwise, a single direction was randomly selected with equal probability. In addition, only the largest strongly connected component of the network was considered, so as to avoid the random walker to become trapped. So, in order to keep the original size of the network as much as possible while incorporating a considerable degree of directionality, we set $r=0.6$.

\subsection{Huffman algorithm}

In digital media, a message can be encoded as a set of organized symbols, which are stored by using a fixed number of bits. For example, texts are formed of symbols which can be represented as characters of $8$ bits. In order to store or transmit messages in an effective way, several lossless compression algorithms have been proposed in the literature~\cite{Salomon:2009:HDC:1708069}. 

The \emph{Huffman code} is a particular data compression algorithm, based on information theory~\cite{Cover:2006:EIT:1146355}. This method generates a dictionary of bit sequences employed to represent each symbol in a message. The compression is achieved by associating shorter bit sequences to more frequent symbols and longer sequences to symbols that appears more rarely in the message. To do so, the Huffman algorithm uses a binary tree, whose leaves represent symbols. Starting from the root, every edge is associated to a bit. Typically, left and right children are associated to the bits 0 and 1, respectively.  The symbol code associated to each leaf node is then obtained by concatenating, from root to leaves, all edge values from the root. Such structure is used as a dictionary to encode the original message. In a similar fashion, the dictionary is used to decode the message. 

\subsection{Network reconstruction}

As the generated time series reaches the receiver, it is used to progressively reconstruct the original network being transmitted.  This can be easily achieved by starting with a disconnected set of symbols and adding each new received edge, defined by a pair of subsequent symbols in the time series, to the network.  We expect the reconstruction to depend on: i) the size of the time series; ii) the considered dynamics; iii) the network topology.  In the case of the considered random walks, perfect reconstruction is expected after a sufficient period of time.

\section{RESULTS AND DISCUSSION}

The main question to be investigated experimentally regards how effectively each of the network models can be recovered from a respectively generated time series, in presence or not of compression.  For every type of network, each of the considered random walks is applied in order to probe the respective topology by visiting, sequentially, the nodes.  A respective time series is generated in the process from which, at each time instant, a reconstruction of the original is obtained by considering the edges and nodes already transmitted.  Therefore, the efficiency of the transmission can be quantified in terms of the time series length (which is proportional to the transmission time) required for reconstruction of 90\% of the original network (measured in terms of number of edges). This critical time is henceforth referred to as $T_{90}$.  The effect of compressing the time series by using the frequency of visits to nodes predicted by the respective degrees, referred to as $T_{90}^C$, is also considered, giving rise to another series of experiments. Furthermore, we also calculated the long term transmission rate, defined as $R_{L} = T^{C}_{L}/ T_{L}$, where $T_{L}$ is a sufficiently large time (a total of one million symbols was used in the reported experiments).  In principle, a combination of topology and dynamics that allows large compression ratio should lead to faster reconstruction. 

\begin{figure*}[!htpb]
  \centering
    \includegraphics[width=\textwidth]{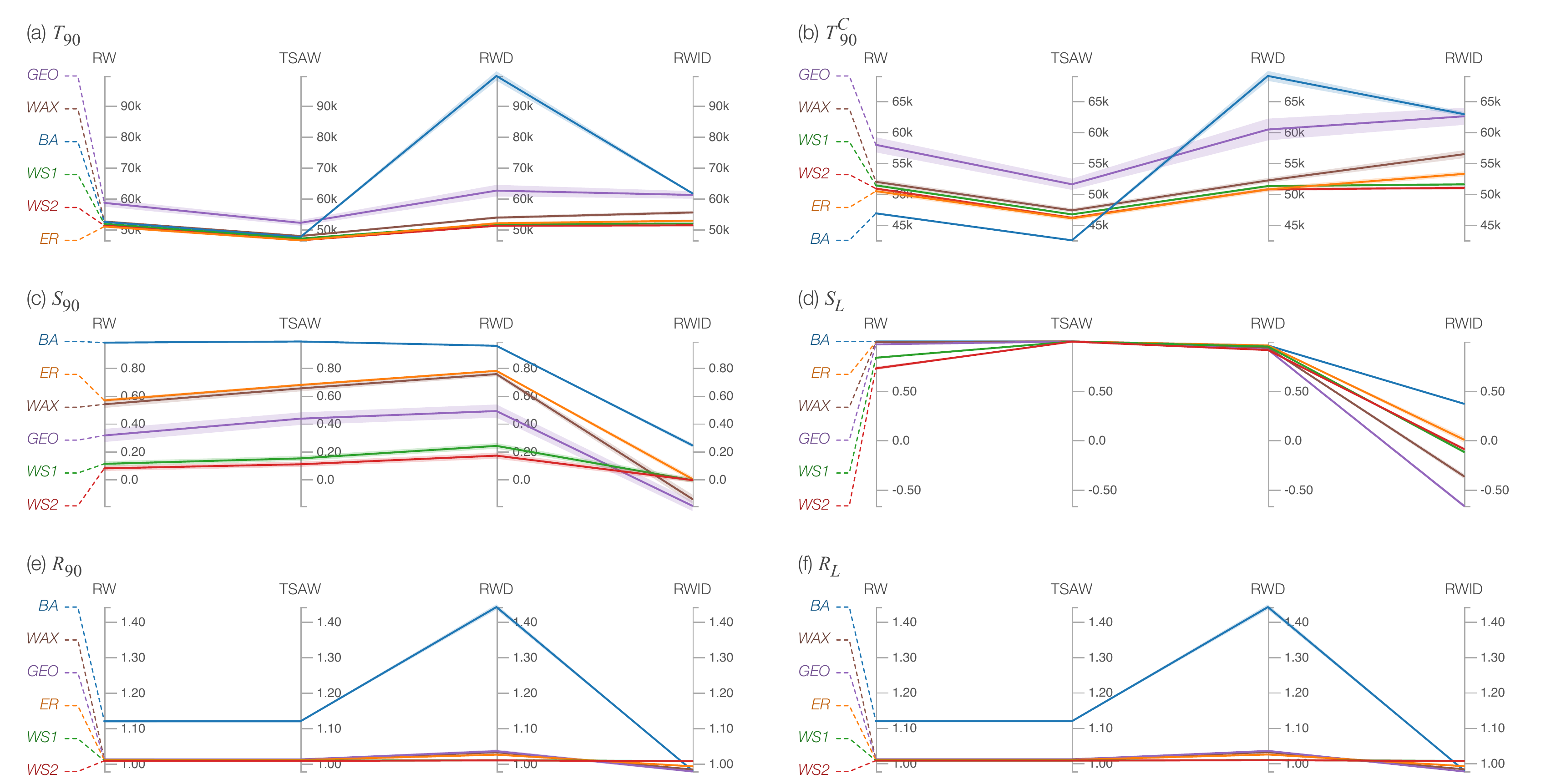}
  \caption{Transmission and compaction average values and standard deviations obtained for \emph{undirected} networks for diverse types of random walks.  (a) Time to transmit 90\% of the network ($T_{90}$); (b) Time to transmit 90\% of the network with Huffman compression ($T_{90}^C$); (c) Steering coefficient attained after exploring 90\% of the network ($S_{90}$); (d) Steering coefficient for longer term exploration ($S_L$); (e) Compression ratio for $90\%$ exploration ($R_{90}$), and (f) Compression ratio obtained after a large number of iterations ($R_L$).}
  \label{f:resUndirected}
\end{figure*}

Fig.~\ref{f:resUndirected} shows the parallel coordinates obtained for the undirected network models.  A total of 30 simulations was performed for each model and dynamics, and 30 network realizations were used for each network model.  The considered networks had approximately 1000 nodes and average degree near 8. The overall best transmission times were obtained for the TSAW random walk.  The BA implied a substantially higher value of $T_{90}$ for the RWD dynamics (Fig.~\ref{f:resUndirected}(a)).  This is probably a consequence of the fact that, in this dynamics, the moving agent tends to alternate between hubs, overlooking nodes with small degree.  As shown in Fig.~\ref{f:resUndirected}(b), similar results were obtained when considering compressed times, $T_{90}^C$, although in this case the values of $T_{90}^C$ for the RWD dynamics in the BA model are not as prominent as those obtained for $T_{90}$.  The results for the WAX and GEO models differed significantly, with the former being transmitted much more effectively.  This result is surprising because both these models share a geographical nature, in the sense that nodes that are spatially close one another tend to be connected.  

Also shown in Fig.~\ref{f:resUndirected} is the steering coefficient ($S_{90}$). Similar values were obtained for the RW, RWD and TSAW dynamics. In the RWID case, the degree is not a good predictor of the frequency of visits, as indicated in Eq.~\ref{eq:longTermProbApprox}, which leads to low $S_{90}$. The WS1 and WS2 models usually led to low steering coefficient values. This probably happens because most nodes in these networks have the same degree, and therefore they also possess similar frequency of visits. The BA model always resulted in the largest steering coefficient.   Such an effect is possibly a consequence of the power-law nature of the degree distribution, in which nodes with small degree tend to be scarcely visited while the opposite happens for nodes with large degree. A more in-depth discussion about the encoding efficiency is presented in Section S1 of the supplementary material, where we compare the compression achieved when estimating the symbol probabilities using the nodes degrees with those achieved when using individual messages to estimate the probabilities.  Remarkably, in most cases we found  that predicting the symbol probabilities from the network (instead of from the sequence of symbols) yielded better compaction.

The long term steering coefficient values ($S_{L}$) are shown in Fig.~\ref{f:resUndirected}(d).  The TSAW dynamics led to the highest $S_L$ values, followed closely by the RWD.  A variety of behavior were observed for RWID, all of them yielding $S_L$ values smaller than those obtained for the other dynamics.

Fig.~\ref{f:resUndirected}(e,f) shows the compression ratio for $90\%$ of network recovery, $R_{90}$, and for long term exploration, $R_{L}$, obtained for the several network models and dynamics.  Similar results were obtained for most cases, except for the BA model in the RW, TSAW and RWD cases, which yielded better compression rates.  Interestingly, the GEO and WAX, which had produced substantially different compression and transmission times in the previous experiment, implied similar compression ratios.

\begin{figure*}[!htpb]
  \centering
    \includegraphics[width=\textwidth]{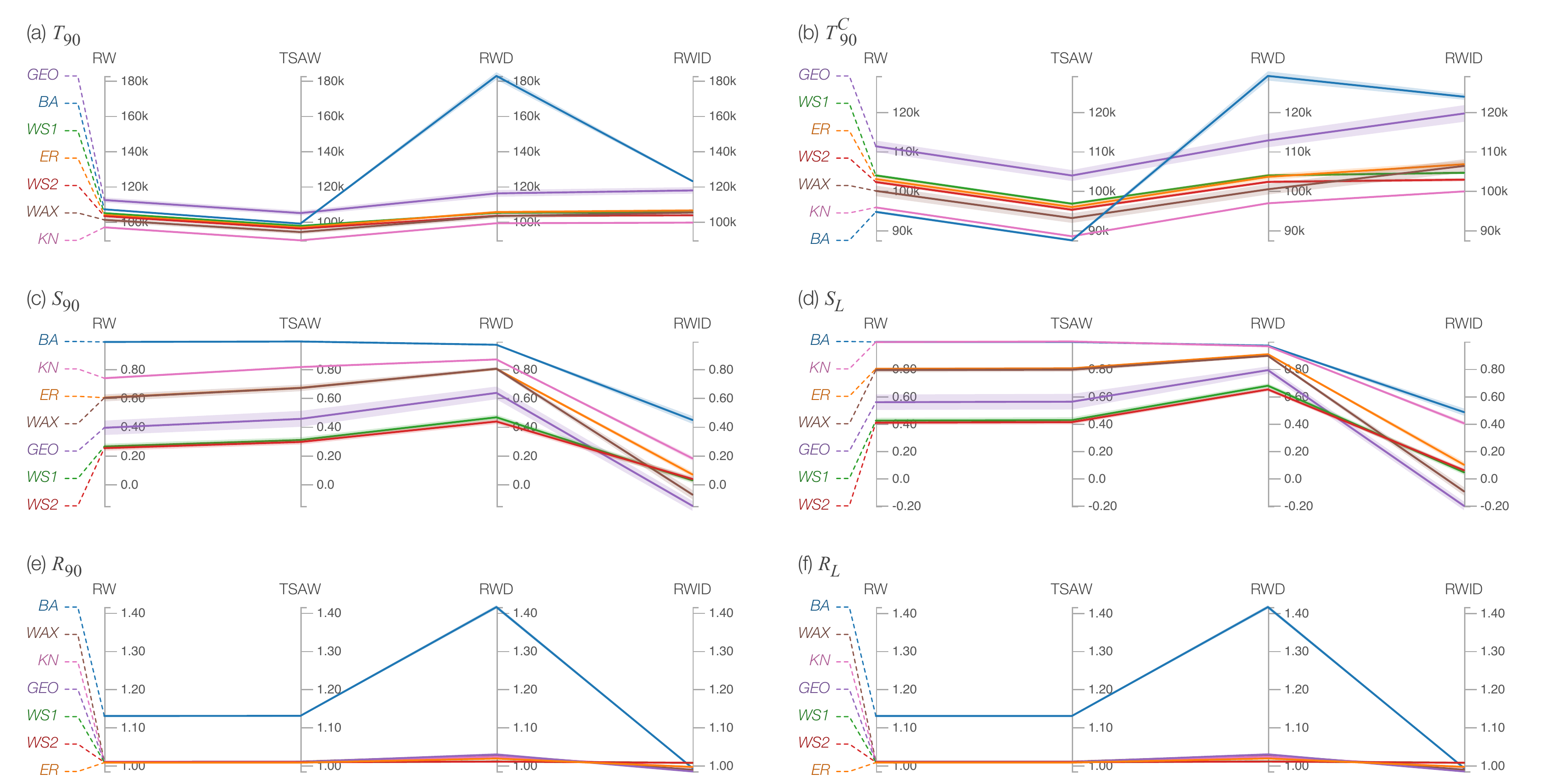}
  \caption{Transmission and compaction average values and standard deviations obtained for \emph{directed} networks for diverse types of random walks.  (a) Time to transmit 90\% of the network ($T_{90}$); (b) Time to transmit 90\% of the network with Huffman compression ($T_{90}^C$); (c) Steering coefficient attained after exploring 90\% of the network ($S_{90}$); (d) Steering coefficient for longer term exploration ($S_L$); (e) Compression ratio for $90\%$ exploration ($R_{90}$), and (f) Compression ratio obtained after a large number of iterations ($R_L$).}
  \label{f:resDirected}
\end{figure*}

The experimental results for the directed cases are shown in Fig.~\ref{f:resDirected}. In this experiment we have the inclusion of the KN model (intrinsically directed), which always led to low reconstruction times in all cases. The better transmission obtained for the KN networks is possibly related to the process of network construction, which can be understood as a kind of random walk. This type of network has two key aspects: (i) for every node, the inward degree ($k_{\mathrm{in}}$) is equal to the outward degree ($k_{\mathrm{out}}$) and (ii) the reciprocity is smaller than the other considered networks. In Section S2 of the supplementary material we compare the transmission times of KN networks with those obtained for networks generated by a directed ER model (thus having reciprocity close to zero) and a configuration model~\cite{Newman:2010:NI:1809753} having the constraint $k_{\mathrm{in}}=k_{\mathrm{out}}$. The results indicate that these two aspects are responsible for the better transmission of KN networks. Returning to Fig.~\ref{f:resDirected}, the results were again similar for $T_{90}$ and $T_{90}^C$, with the exception of the BA model. This model has large $T_{90}$ for the RWD dynamics. Overall, the values of $S_{90}$ showed similar trends as in the undirected case. The long term steering coefficients, $S_L$, resulted smaller than for the undirected cases, with exception of the BA model, which was similar to that case.

The compression ratios $R_{90}$ and $R_L$ obtained for the directed networks are shown in Fig.~\ref{f:resDirected}(e,f). These results are generally similar to those obtained for undirected networks.

\section{CONCLUSIONS}

The areas of information theory and complex networks have been developed in a mostly independent way.  However, as argued in the present work, these two areas present several shared and complementary elements which, when integrated, can be used to model, characterize and analyze a broad range of important real-world problems ranging from spoken/written language to DNA sequences. A formal framework, leading to a potentially new area, has been reported, involving the transmission of an original network, by using a sampling dynamics such as random walks, which produces a sequence of symbols or time series that can be used by a receiver to reconstruct the original network.  More effective transmission demands compaction of the time series which, we argue, is directly related to the topology of the original network.  We also show that the critical issue of compaction is directly related to one of the central paradigms in network science, namely the relationship between topology and dynamics, more specifically regarding the ability to predict the frequency of symbols from the very topology of the original network.  Interestingly, the quality of such a prediction depends on the interplay between the topology of the original network and the adopted dynamics. 

Interestingly, the proposed basic framework can be directly extended to model more sophisticate systems involving several pairs of transmitter-receivers, which are themselves organized as complex networks.  In addition to proposing the systematic integration between network science and information theory, we also illustrated a typical problem that can be tackled in this area, namely the efficiency of transmission of several types of networks by using different kinds of random walks.  A number of interesting results has been reported.  First, we confirmed that different network topologies and dynamics can lead, irrespectively of compaction, to rather distinct performances.  Interestingly, the BA model exhibited a markedly distinct behavior, oscillating between the best and worst performances, depending on the probing random walk.  On the other hand, the KN model, in almost all cases,  led to the best performance in the case of directed networks.  In addition, the two adopted geographical networks, namely WAX and GEO, despite their seemingly analogous spatial organization, yielded rather different results in the case of undirected networks. It is particularly interesting to observe that the BA model, which is topologically very complex (non-uniform degree distribution), led to the best overall compaction in most cases, except the RWID.  This is because the power law degree distribution implies in asymmetric distribution of frequency of symbols, and therefore, more effective Huffman coding. 

Given the generality of the proposed framework with respect to theoretical and practical aspects, the prospects for future works are particularly wide and a more complete list of possibilities would be beyond the scope of this work.  Some particularly promising application venues include the modeling of opinion spreading, syllabuses planning, and language evolution.  Also, it would be interesting to consider noisy transmission, as well as higher order statistical coding of symbols.  Regarding network topology, it would be particularly interesting to investigate how modular structure can impact the transmission.  Other types of dynamics can be also considered, especially those related to neuronal signal propagation.

\section*{Acknowledgements}
The authors acknowledge financial support from Capes-Brazil, S\~ao Paulo Research Foundation (FAPESP) (grant no. 2016/19069-9, 2015/08003-4, 2015/18942-8, 2014/20830-0 and 2011/50761-2), CNPq-Brazil (grant no. 307333/2013-2) and NAP-PRP-USP.

\bibliography{referencias}

\newpage

\renewcommand\thefigure{S\arabic{figure}}    
\setcounter{figure}{0}  

\section*{Supplementary material}
\subsection{S1. Efficiency of network encoding}

\begin{figure*}[!htpb]
  \centering
    \includegraphics[width=\textwidth]{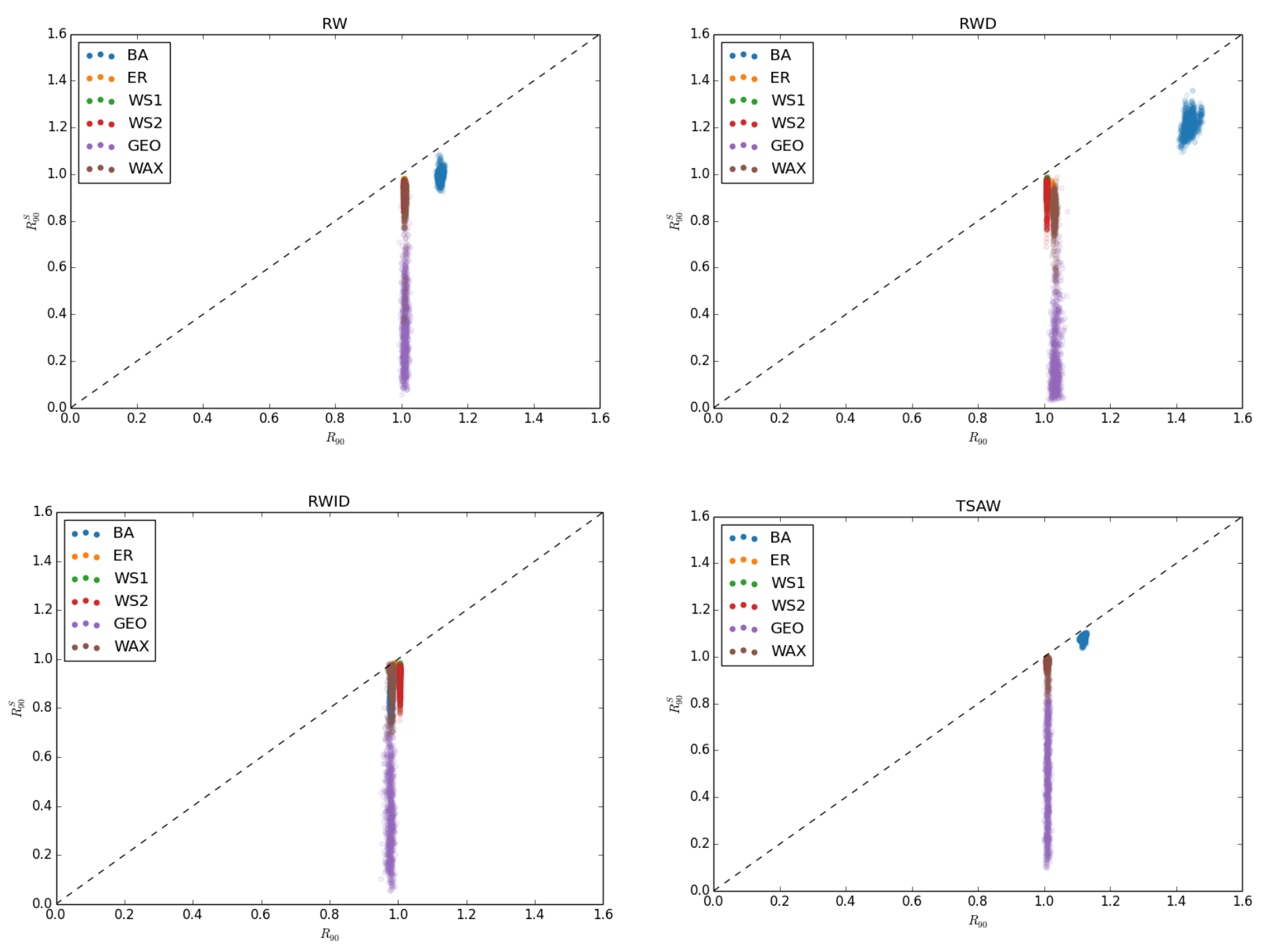}
  \caption{Comparison between compression ratios obtained when using the networks degrees to generate the dictionary ($R_{90}$) and when one of the transmitted messages is used for the dictionary creation ($R_{90}^S$). The plots are respective to the (a) traditional, (b) degree-biased (c) inverse degree-biased and (d) self-avoiding random walk dynamics taking place on undirected networks.}
  \label{f:resHuffvsDegUndirected}
\end{figure*}

As noted in the main text, the steering coefficient ($S$) indicates how well the symbol probabilities of the transmitted message can be predicted from some network topological property. A combination of network and dynamics possessing large $S$ should lead to a good prediction of the statistics of the message. In order to verify if this is indeed the case, we compared the compression ratios achieved when exploring 90\% of the network ($R_{90}$) with those obtained when using one of the transmitted messages to estimate the symbol probabilities of a whole set of messages being transmitted ($R_{90}^S$). The results for the undirected networks are shown in Fig.~\ref{f:resHuffvsDegUndirected}. It is clear that estimating the probabilities using the topology of the system always resulted in better compression ratios than the single message approach. Also, observe that the GEO model displays a large variance of $R_{90}^S$ values. This is caused by the large diameter of this network, which leads to substantial parts not being visited by the walker when the symbol probabilities are being estimated. This leads to low compression ratios for messages starting around nodes that were not visited during the probability estimation, while larger compression ratios are obtained for messages starting closer to the original message used for constructing the symbol dictionary.

\begin{figure*}[!htpb]
  \centering
    \includegraphics[width=\textwidth]{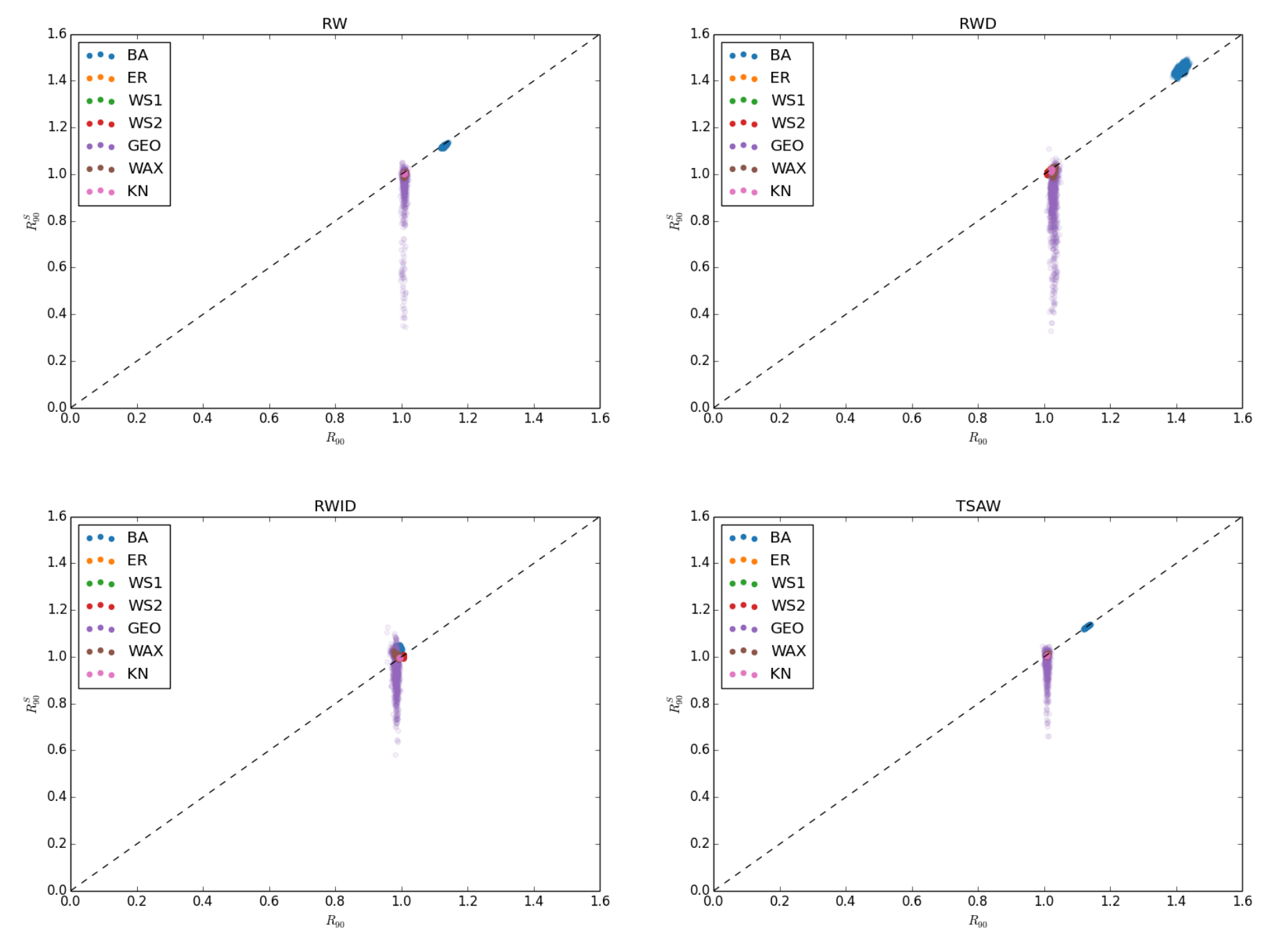}
  \caption{Comparison between compression ratios obtained when using the networks degrees to generate the dictionary ($R_{90}$) and when one of the transmitted messages is used for the dictionary creation ($R_{90}^S$). The plots are respective to the (a) traditional, (b) degree-biased (c) inverse degree-biased and (d) self-avoiding random walk dynamics taking place on directed networks.}
  \label{f:resHuffvsDegDirected}
\end{figure*}

The comparison between $R_{90}$ and $R_{90}^S$ obtained for directed networks is shown in Fig.~\ref{f:resHuffvsDegDirected}. In this case, similar values were obtained for both quantities. This is likely due to the slower random walk exploration of edges allowed by the directed networks. This, in turn, leads to better estimation of the node probabilities, since the walker is more likely to reach nodes that would not be visited otherwise if the network were undirected.

\subsection{S2. Better exploration times of knitted networks}

\begin{figure*}[!htpb]
  \centering
    \includegraphics[width=\textwidth]{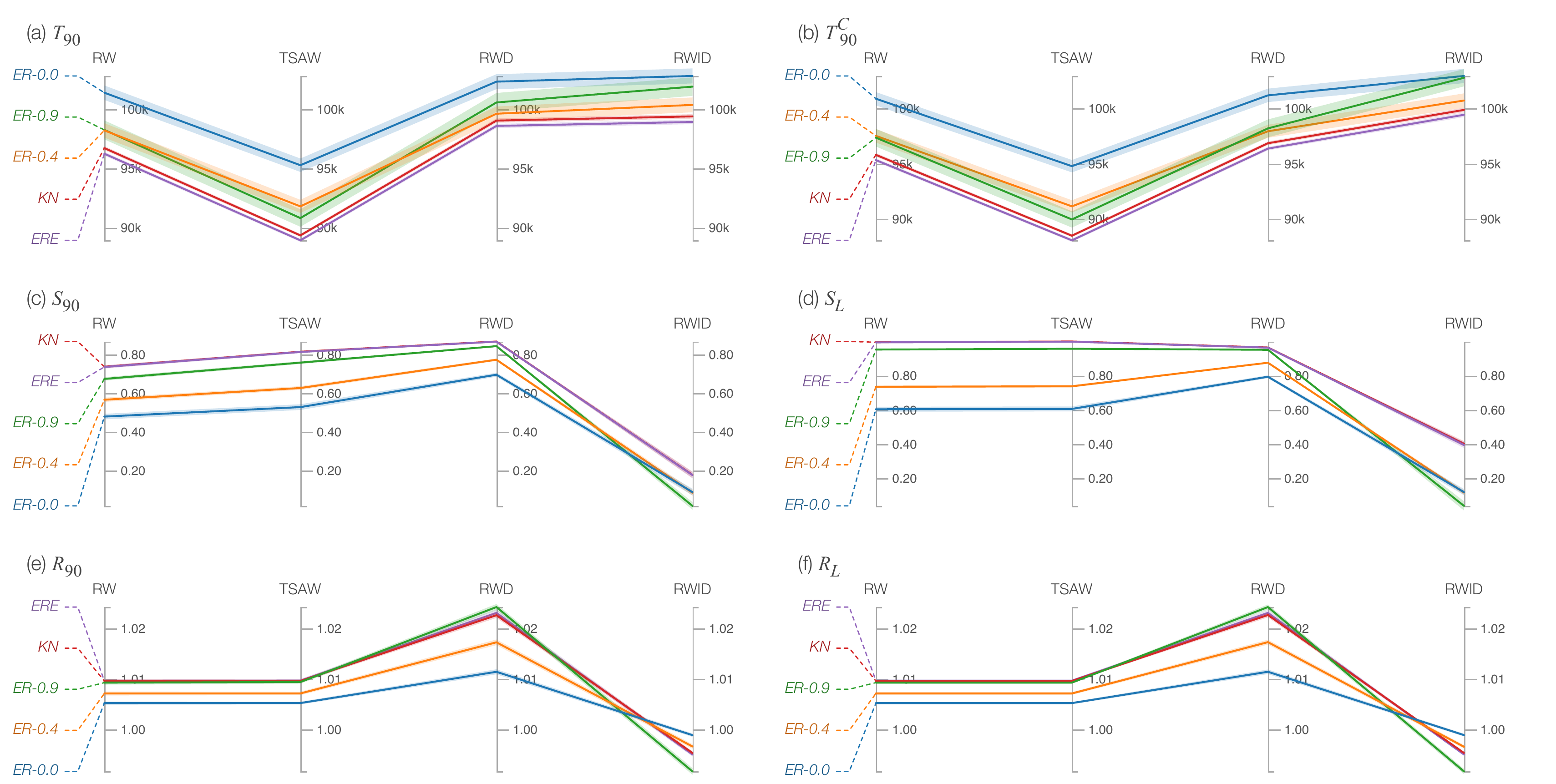}
  \caption{Transmission and compaction average values and standard deviations obtained for variations of the ER model and for KN networks. Shown are the values for ER networks having distinct reciprocities (indicated as $ER-0.0, ER-0.4$ and $ER-0.9$), and networks where $k_{\mathrm{in}}=k_{\mathrm{out}}$ (ERE). (a) Time to transmit 90\% of the network ($T_{90}$); (b) Time to transmit 90\% of the network with Huffman compression ($T_{90}^C$); (c) Steering coefficient attained after exploring 90\% of the network ($S_{90}$); (d) Steering coefficient for longer term exploration ($S_L$); (e) Compression ratio for $90\%$ exploration ($R_{90}$), and (f) Compression ratio obtained after a large number of iterations ($R_L$).}
  \label{f:kn_supplementary}
\end{figure*}

In order to understand why the simulations executed in Knitted networks (KN) resulted in better exploration times, we compare the results of this model with a set of ER networks having distinct reciprocity values and different restrictions regarding the in-degree ($k_{\mathrm{in}}$) and out-degree ($k_{\mathrm{out}}$) of each node. First, networks were created with a range of reciprocity values, which are approximately $0.0$, $0.4$ and $0.9$. The methodology used for changing the reciprocity of the networks is described in Section II of the main text. The four random walk dynamics considered in the main text were applied to the networks. The obtained exploration times ($T_{90}$ and $T_{90}^C$), steering coefficients ($S_{90}$ and $S_L$) and compression ratios ($R_{90}$ and $R_L$) are shown in Fig.~\ref{f:kn_supplementary}. Note that $T_{90}$ and $T_{90}^C$ increase with the reciprocity. Furthermore, Fig.~\ref{f:kn_supplementary} also shows the results for ER networks built so that the reciprocity is close to zero and $k_{\mathrm{in}}$ is equal to $k_{\mathrm{out}}$ for each node (ERE). For all the considered random walks, the exploration times $T_{90}$ and $T_{90}^C$ of the KN networks were mostly similar to those obtained for ERE networks. Therefore, we believe that the main characteristics leading to an efficient exploration in KN networks are the low reciprocity and the fact that, by construction, $k_{\mathrm{in}}$ is equal to $k_{\mathrm{out}}$ for almost all nodes.

\end{document}